# Superconducting Magnetic Bearings Simulation using an H-formulation Finite Element Model

Loïc Quéval[1], Kun Liu[2], Wenjiao Yang[2], Víctor M. R. Zermeño[3], Guangtong Ma[2*]

[1]*Group of electrical engineering - Paris (GeePs), CNRS UMR 8507, CentraleSupélec, UPSud, UPMC, Gif-sur-Yvette, France. Phone: +33-169851534, e-mail address: loic.queval@geeps.centralesupelec.fr*

[2]*Applied Superconductivity Laboratory (ASCLab), State Key Laboratory of Traction Power, Southwest Jiaotong University, Chengdu, Sichuan 610031, China. Phone: +86-28-87603310, e-mail address: kunliu_swjtu@163.com, motor_yang@163.com, gtma@swjtu.edu.cn*

[3]*Karlsruhe Institute of Technology, Hermann-von-Helmholtz Platz 1, 76344 Eggenstein-Leopoldshafen, Germany Phone: +49-72160828582, e-mail address: victor.zermeno@kit.edu*

*\* Author to whom correspondence should be addressed*

**Abstract**

The modeling of superconducting magnetic bearing (SMB) is of great significance for predicting and optimizing its levitation performance before construction. Although lots of efforts have been made in this area, it still remains some space for improvements. Thus the goal of this work is to report a flexible, fast and trustworthy H-formulation finite element model. First the methodology for modeling and calibrating both bulk-type and stack-type SMB is summarized. Then its effectiveness for simulating SMBs in 2-D, 2-D axisymmetric and 3-D is evaluated by comparison with measurements. In particular, original solutions to overcome several obstacles are given: clarification of the calibration procedure for stack-type and bulk-type SMBs, details on the experimental protocol to obtain reproducible measurements, validation of the 2-D model for a stack-type SMB modeling the tapes real thickness, implementation of a 2-D axisymmetric SMB model, implementation of a 3-D SMB model, extensive validation of the models by comparison with experimental results for field cooling and zero field cooling, for both vertical and lateral movements. The accuracy of the model being proved, it has now a strong potential for speeding up the development of numerous applications including maglev vehicles, magnetic launchers, flywheel energy storage systems, motor bearings and cosmic microwave background polarimeters.

## I. Introduction

The relative movement between a permanent magnet (PM) and a high temperature superconductor (HTS) can induce supercurrents in the HTS. By interacting with the PM static magnetic field, these supercurrents produce a force that can be attractive or repulsive depending on the arrangement and on the operating conditions. It can even provide passive stable levitation. This unique feature motivated the development of superconducting magnetic bearings (SMBs) [1-3]. They have been customized for numerous applications, including maglev vehicles [4-6], magnetic launchers [7, 8], flywheel energy storage systems [9-16], motors [17], and cosmic microwave background polarimeters [18-20].

There are various analytical and numerical models that are able to predict, more or less accurately, the maglev performances of SMBs. A detailed review is provided by Navau *et al.* in [21]. Among them, finite element (FE) models using various formulations are being intensively developed. The formulations are named after the state variables to be solved: A-V-formulation for the magnetic potential vector and the electric potential, T-Ω-formulation for the current potential vector and the magnetic potential, E-formulation for the electric field and H-formulation for the magnetic field. The critical state model (CSM) [22] or the E-J power law model [23] is then commonly used together with one of these formulations to model the nonlinear resistivity of the superconductor. A summary of the formulation used by independent groups to model SMBs with homemade FE codes and free/nonfree FE softwares is proposed in Table I.

TABLE I
SMB FINITE ELEMENT MODELS

|   |   | 2-D | 2-D axi | 3-D |
|---|---|---|---|---|
| A-V | Homemade | Hofmann et al. [24]<br>Dias et al. [25-27]<br>Ma et al. [28-30] | Sugiura et al. [31]<br>Takeda et al. [32]<br>Chun et al. [33]<br>Ruiz-Alonso et al. [34]<br>Wang et al. [35]<br>Sotelo et al. [36] | Ueda et al. [38] |
|   | Software | - | Li et al. [37] | Hauser [39] |
| T-$\Omega$ | Homemade | Zheng et al. [40]<br>Zhang et al. [41] | Gou et al. [42] | Uesaka et al. [43, 44]<br>Tsuchimoto et al. [45]<br>Tsuda et al. [46-48]<br>Ma et al. [49, 50]<br>Pratap et al. [51] |
|   | Software | - | - | - |
| E | Homemade | - | - | - |
|   | Software | - | - | - |
| H | Homemade | Lu et al. [52] | - | Lu et al. [56]<br>Yu et al. [57] |
|   | Software | Sass et al. [53]<br>Quéval et al. [54]<br>*this work* | Patel et al. [55]<br>*this work* | Patel et al. [55]<br>Quéval et al. [54]<br>*this work* |

All these models have their own features and limitations. Focusing on the H-formulation, important efforts have been made to simulate SMBs using homemade codes. Lu et al. wrote a FE code in FORTRAN to estimate the levitation force between a PM and an HTS bulk in 2-D [52]. This is probably an evolution of the code reported in [56] for the 3-D simulation of a cylindrical HTS bulk over a PM guideway. In those articles, the field of the moving PM, obtained analytically, was applied as a time dependent Dirichlet boundary condition on the outer boundary of a model including only the HTS domain and a thin air domain. But it is not clear if the self-field of the HTS bulk was included. The model and its extensions to other PM guideways geometries, field cooling and lateral movements [58-60] provided interesting guidelines but the authors provided no convincing experimental validation of it. Yu et al. implemented a similar 3-D model to analyze a SMB made of a cylindrical PM and a cylindrical HTS bulk [57]. A substantial effort was made there to experimentally validate the model for both zero field cooling and field cooling, but only for vertical displacements. Surprisingly, the simulated levitation force did not go back to zero when the gap increased. And the levitation force loop proved difficult to reproduce for the field cooling case.

FE softwares have also been employed to simulate SMBs using the H-formulation. Actually the groups listed in Table I all used Comsol Multiphysics [61], either with the magnetic field formulation (mfh) physic available in the AC/DC module, or by manually implementing the partial differential equations (PDEs) with the PDE module. Sass et al. developed a 2-D model [53] to obtain the levitation force between a PM and an YBCO bulk or stacks of YBCO tapes. The field of the PM was obtained using analytical equations. To model the movement, the field generated by the PM was applied as a time dependent Dirichlet boundary condition on a boundary close to the HTS domain. To reduce the computing time, a symmetry axis was used, restricting the movement to vertical displacements. To model the stacks, an anisotropic homogenized model was adopted [62]. The agreement with measurements for field cooling and zero field cooling was good. A similar model was developed by Quéval et al. [54] to include the PM assembly real geometry and the iron nonlinearity. To do so, the field of the PM assembly was obtained using a magnetostatic FEM. Besides, the model was able to deal with any relative movement, making it possible to optimize the SMB on a realistic displacement sequence. A similar 3-D model was mentioned in [54] but without details about its implementation. Patel et al. introduced a 2-D axisymmetric H-formulation FEM in [55] to estimate the levitation force between a PM and stacks of YBCO tapes. The PM was modeled by a thin current domain approximating the ideal equivalent 2-D axisymmetric current sheet. To model the movement, this thin domain was moved along the z-direction by defining it with a time and space dependent current density. With this modeling strategy, the boundary conditions are fixed but many elements are required to mesh the "moving" PM assembly thus limiting the applicability of the model to simple geometries. To model the stacks, an isotropic homogenized model was used. The simulated levitation force, limited to the first magnetization, was compared with measurements from 20 K to 77 K in field cooling condition only. The agreement for a SMB with a rolled stack was fair at 20 K and reasonable at 77 K [55]. For a SMB with a stack of annuli [63], the agreement was good. Similarly, a 3-D model was built to study the current pattern for the SMB with the rolled stack with limited discussion and validation [55].

The motivation behind this work is to develop flexible, fast and trustworthy H-formulation FE models able to predict the maglev performances of SMBs in 2-D, 2-D axisymmetric and 3-D configurations. Key advancement with respect to previous models include: clarification of the calibration procedure for stack-type and bulk-type SMBs, details on the experimental protocol to obtain reproducible measurements, validation of the 2-D model for a stack-type SMB considering the tapes real thickness, implementation of a 2-D axisymmetric SMB model, implementation of a 3-D SMB model, extensive validation of the models by comparison with experimental results for field cooling and zero field cooling, for both vertical and lateral movements. The test cases reported here have been selected to serve as benchmarks, with the hope to help focus the effort of the numerical modelling community towards the most relevant approaches [64].

## II. Superconducting magnetic bearing model

The SMB model is built by unidirectional coupling between the PM assembly model and the HTS assembly model. The coupling is done by applying the sum of the external field $\mathbf{H}_{ext}$ and the self-field $\mathbf{H}_{self}$ on the outer boundaries $\Gamma$ of the HTS assembly model (Figure 1).

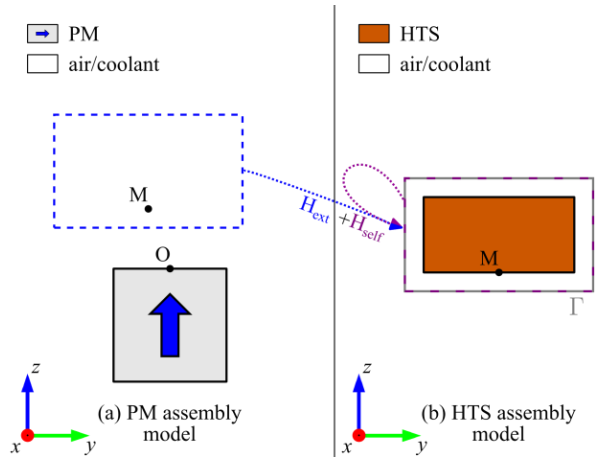

Figure 1 – Modeling approach. (a) PM assembly model. O is the origin of the coordinate system. The outer boundary is not shown. The time-dependent coordinates of M in the PM assembly reference frame describe the relative movement of the HTS and PM assemblies. (b) HTS assembly model. M is the origin of the coordinate system.

1) PM assembly model

The PM assembly is an arrangement of any number of PMs and ferromagnetic pieces surrounded by air (or any coolant). For simple geometries, analytical formulas could be used [65, 53]. But it is modeled here using a magnetostatic A-formulation FE model. This allows us to include the iron nonlinear B-H curve and to consider complex PM assembly geometries [54].

2) HTS assembly model

The HTS assembly is an arrangement of any number of normal and superconducting pieces (bulks or conductors) surrounded by air (or any coolant). It is assumed that the materials are non-magnetic. To mathematically model the HTS assembly, the H-formulation is used [66, 67],

$$\nabla \times \rho \nabla \times \mathbf{H} = -\mu_0 \frac{\partial \mathbf{H}}{\partial t} \text{ in } \Omega \tag{1}$$

$$\mathbf{H} = \mathbf{H}_{self} + \mathbf{H}_{ext} \text{ on } \Gamma \tag{2}$$
$$\mathbf{H}(t_0) = \mathbf{H}_0 | \nabla \cdot (\mu_0 \mathbf{H}_0) = 0 \tag{3}$$

where $\mathbf{H}$ is the magnetic field strength, $\rho$ is the material resistivity, $\mu_0$ is the vacuum magnetic permeability, $\Omega$ is the computational domain and $\Gamma$ is the outer domain boundary. Neumann boundary conditions are used for inner boundaries. On $\Gamma$, Dirichlet boundary conditions are used to impose the self-field $\mathbf{H}_{self}$ (the one created by the supercurrent) and the external field $\mathbf{H}_{ext}$ (the one created by the PM assembly). The current density $\mathbf{J}$, the electric field $\mathbf{E}$ and the magnetic flux density $\mathbf{B}$ can be obtained from $\mathbf{H}$ using,

$$\mathbf{J} = \nabla \times \mathbf{H} \tag{4}$$
$$\mathbf{E} = \rho \mathbf{J} \tag{5}$$
$$\mathbf{B} = \mu_0 \mathbf{H} \tag{6}$$

The resistivity $\rho_{sc}$ of the HTS is represented by a power law,

$$\rho_{sc}(|\mathbf{J}|, \mathbf{B}) = \frac{E_c}{J_c(\mathbf{B})} \left| \frac{\mathbf{J}}{J_c(\mathbf{B})} \right|^{n-1} \tag{7}$$

where $J_c$ is the field dependent local critical current density, $E_c$ is the critical current criterion and $n$ is a material parameter. To impose a transport current in a conductor, an integral constraint on the current density can be used

$$I_{tr}(t) = \int_{\Omega_c} \mathbf{J} \cdot d\mathbf{s} \tag{8}$$

where $\Omega_c$ is the conductor cross section and $d\mathbf{s}$ is the differential cross-sectional area vector. For the finite element discretization, we use linear edge elements [66].

The external field $\mathbf{H}_{ext}$ is obtained from the PM assembly model. The static magnetic field generated by the PM assembly $\mathbf{H}_{PM}$ needs to be modified to take the relative movement into account. This is done by

$$\mathbf{H}_{ext}(x, y, z, t) = T_t \, \mathbf{H}_{PM}(x, y, z) \tag{9}$$

where $T_t$ is the translation operator that describes the time-dependent position of the HTS assembly in the PM assembly reference frame.

The HTS is said to be "field cooled" (FC) when the cooling is achieved close to the PM assembly, and "zero field cooled" (ZFC) when the PM is far enough so that the applied field is negligible. We assume that during the cooling all the flux is pinned [68] and that no macroscopic currents are induced in the HTS [69]. This is experimentally validated by the fact that the forces after cooling but before any movement are null [26]. To simulate the FC case, we can therefore disregard $\mathbf{H}_{self}$ and set $\mathbf{H}_0 = T_{t_0} \mathbf{H}_{PM}(x, y, z)$. By doing so, (3) is respected because the divergence of the field generated by the PM is zero. Note that we implicitly make here the hypothesis that the field generated by the supercurrent does not influence the PM's remanent field.

The self-field $\mathbf{H}_{self}$ is obtained from the HTS assembly model at each time step by numerical integration of the Biot-Savart law. The consideration of $\mathbf{H}_{self}$ is required to make the problem self-consistent since the air/coolant layer around the HTS domain is slim. Indeed $\mathbf{H}_{ext}$ is applied on a boundary that is close to the HTS domain.

The force $\mathbf{F}$ between the PM assembly and the HTS assembly is obtained with

$$\mathbf{F} = \int_{\Omega_{sc}} \mathbf{J} \times \mathbf{B} \, ds \tag{10}$$

where $\Omega_{sc}$ is the HTS assembly cross section and $ds$ the differential cross-sectional area.

## III. Measurements

The force measurements were carried out using a test rig developed at ASCLab (Figure 2). The 3-D relative motion is obtained by three step motors and screw rods. The 3-D position is recorded by three linear displacement sensors. The 3-D force is measured by a 3-D load cell. The time, the 3-D position and the 3-D force are recorded at 1 kHz. The measured data presented here corresponds to a 500 points moving average.

The PM assembly is at room temperature while the HTS assembly is at liquid nitrogen temperature. A 1 mm sheet of aerogel paper CT200-Z is used to thermally insulate the PM and avoid a shift of its remanent flux density with the temperature during the measurement [70]. The *z*-direction force recorded by the load cell includes the weight of the HTS assembly: therefore the initial force (i.e. the weight) was subtracted from the measurements to remove any force not produced by the supercurrent in the measured data presented here. The liquid nitrogen container is mounted so that its weight is not measured by the load cell.

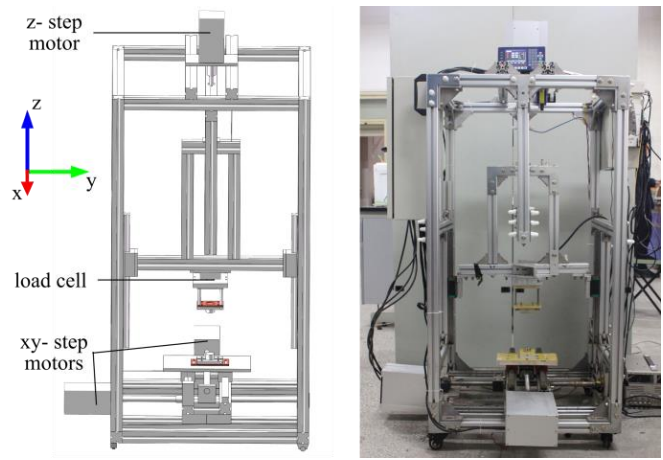

Figure 2 – Test rig used to measure the 3-D forces of the SMBs. The PM assembly was fixed to the base moving in the *xy*-direction. The HTS assembly was fixed to the 3-D load cell moving in the *z*-direction.

## IV. 2D case: linear SMB
### 1) Geometry

The linear SMB and the coordinate system adopted in this section are shown in Figure 3. The PM assembly is made of cuboidal Nd-Fe-B permanent magnets and iron slabs arranged in flux concentration. The HTS assembly is a stack of 120 YBCO tapes (SuperPower SCS12050-AP). The HTS assembly can only move along the *yz*-plane ($x_M(t) = 0$).

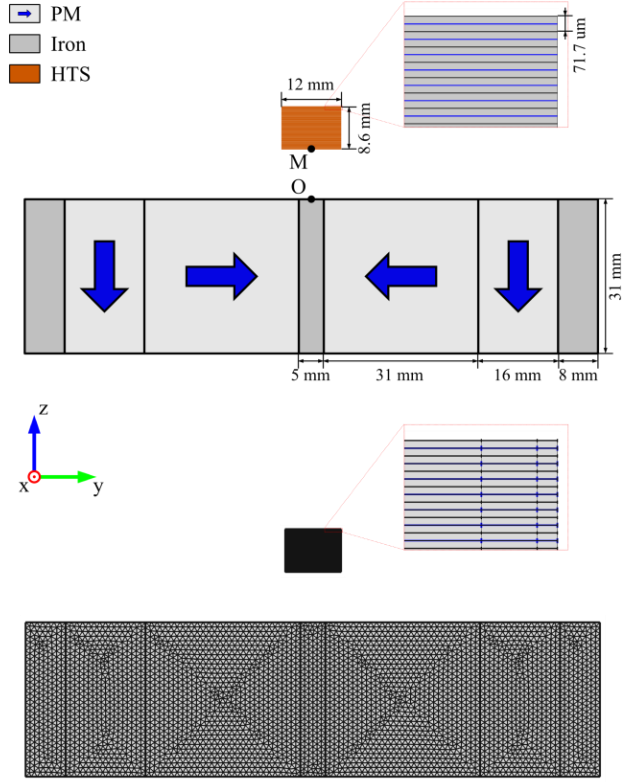

Figure 3 – SMB geometry and mesh for the 2-D case. The point O is located at the center of the PM assembly top surface. The point M is located at the center of the stack bottom surface. The dimension of the PM and HTS assemblies in the *x*-direction are 240 mm and 100 mm, respectively. The arrows indicate the PM magnetization direction. The mesh of the air/coolant is not shown. Inset: zoom on the stack; the blue line show the superconductor layer.

2) Sequences

In this section, we consider three displacement sequences. They are described by the successive positions of $M(y_M, z_M)$ relative to O (in millimeters). The first position of each sequence is the cooling position. The moving speed is 1 mm/s representing a quasistatic process.

- ZFC100: $(y_M, z_M) = \{(0,100), (0,6), (0,100)\}$
- FC25: $(y_M, z_M) = \{(0,25), (0,6), (0,25)\}$
- FC25_LD: $(y_M, z_M) = \{(0,25), (0,6), (6,6), (-6,6), (6,6)\}$

3) Modeling

Equations (1)-(10) are implemented in COMSOL Multiphysics 4.3a PDE mode application in a 2-D space. More details about such implementation can be found in [71] for example. The HTS assembly is a stack of YBCO tapes: we model only the superconducting layers taking their real thickness into account. Each tape has a net current enforced to zero by means of an integral constraint. An anisotropic Kim-like model [72] is used to describe the dependence of the critical current density on the magnetic field,

$$J_c(\mathbf{B}) = \frac{J_{c0}}{\left(1 + \frac{\sqrt{k^2 B_{//}^2 + B_\perp^2}}{B_0}\right)^\alpha} \quad (11)$$

where $B_{//}$ and $B_\perp$ are the field components parallel and perpendicular to the tape, respectively. $J_{c0}$, $B_0$, $k$ and α are material parameters. Equation (11) provides a reasonable description of the anisotropic behavior of HTS coated conductors (without artificial pinning) [73]. To mesh the superconducting layer, we use a mapped mesh [74] with 10 elements distributed symmetrically following an arithmetic sequence in the width and 1 element in the thickness. Such mesh proved to be a good compromise between speed and accuracy. The outer boundary of the HTS assembly model Γ is located at a distance of 1.5 mm from the HTS stack. This is less than the minimum levitation gap so that the coupling boundary is always inside the air gap.

From (9) in 2-D, with the conventions of Fig. 2, the expression for $\mathbf{H}_{ext}$ becomes,
$$\mathbf{H}_{ext}(y, z, t) = \mathbf{H}_{PM}(y + y_M(t), z + z_M(t)) \tag{12}$$

where $(y_M, z_M)$ is the time-dependent position of the HTS assembly relative to O. $\mathbf{H}_{self}$ is obtained by 2-D integration of the Biot-Savart law,
$$H_{self,y}(y, z, t) = \frac{1}{2\pi} \iint_{\Omega_{sc}} \frac{-J_x(y', z', t) \cdot (z - z')}{(y - y')^2 + (z - z')^2} dy'dz' \tag{13}$$
$$H_{self,z}(y, z, t) = \frac{1}{2\pi} \iint_{\Omega_{sc}} \frac{J_x(y', z', t) \cdot (y - y')}{(y - y')^2 + (z - z')^2} dy'dz' \tag{14}$$

where $\Omega_{sc}$ is the HTS assembly domain. This completes and corrects [53].

4) Model calibration

To calibrate the PM assembly model, we need to know the B-H curve of the iron and the remanent flux density $B_r$ of the PM. The assumed B-H curve is given in appendix. To obtain the remanent flux density $B_r$ of the PM, we measured the magnetic flux density at several distances above the PM. By a trial-and-error process, we obtained $B_r$ that minimize the difference between the measured data and the PM assembly model (Figure 4). To calibrate the HTS assembly model, we need to get the values of 5 parameters: $J_{c0}$, $n$, $B_0$, $k$ and α. To obtain $J_{c0}$, it is a common practice to use the maximum levitation force obtained for a zero field cooling sequence [43, 56, 26, 50]. The procedure used here is different. $J_{c0}$ and $n$ are obtained by fitting the power law to the measured current-voltage curve of a short sample of the same conductor. The measurement was made at 77 K using the four probe method. The other HTS tape parameters $B_0$, $k$ and α are obtained by trial-and-error so that the simulated maximum levitation force during the ZFC100 sequence is equal to the measured value (Figure 5). The procedure used here is applicable for any stack-type SMB with the advantage that only 3 parameters are obtained by trial-and-error. The parameters of the 2-D case are summarized in Table II.

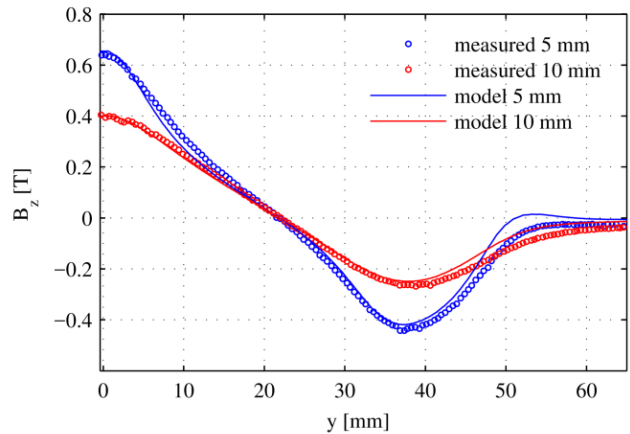

Figure 4 – 2-D model calibration: magnetic flux density at 2 mm and 5 mm above the PM.

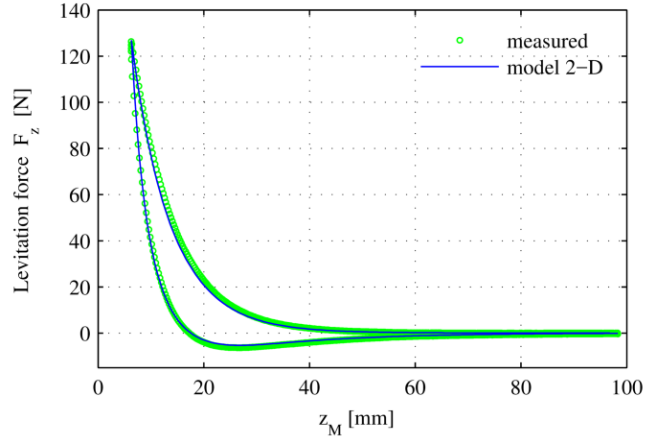

Figure 5 – 2-D model calibration: levitation force for the ZFC100 sequence.

TABLE II
PARAMETERS FOR THE 2-D CASE

| Symbol | Quantity | Value |
| --- | --- | --- |
| $B_r$ | PM remanent flux density | 1.12 T (side) 0.975 T (middle) |
| $E_c$ | Critical current criterion | $1\cdot10^{-4}$ V/m |
| $n$ | HTS parameter | 31 |
| $J_{c0}$ | HTS parameter | $3.225\cdot10^{10}$ A/m$^2$ |
| $B_0$ | HTS parameter | 0.0525 T |
| $k$ | HTS parameter | 0.256 |
| $\alpha$ | HTS parameter | 0.58 |
| $\rho_{air}$ | Air resistivity | 1 Ω·m [75] |
| $\mu_0$ | Air/HTS permeability | $4\pi\cdot10^{-7}$ H/m |

5) Model validation

To validate the 2-D axisymmetric model, we consider the FC25 and FC25_LD sequences. The force calculated with the 2-D model is in good agreement with the measured force (Figure 6 and Figure 7). This validates the modeling approach adopted. Similar simulations were performed for the stack-type SMB analyzed in [53] giving similar agreements (not reported here). Note the four main differences between this model and the one previously reported in [53]. First, we obtain $\mathbf{H}_{ext}$ using a magnetostatic FEM while Sass *et al.* used analytical formulas. Second, we model the stack with tapes real thickness whereas Sass *et al.* used an anisotropic homogenized bulk model [62]. Third, we use an elliptical $J_c(\mathbf{B})$ model whereas Sass *et al.* used an exponential model. Fourth, Sass *et al.* considered only a vertical displacement while we include here both vertical and lateral displacements.

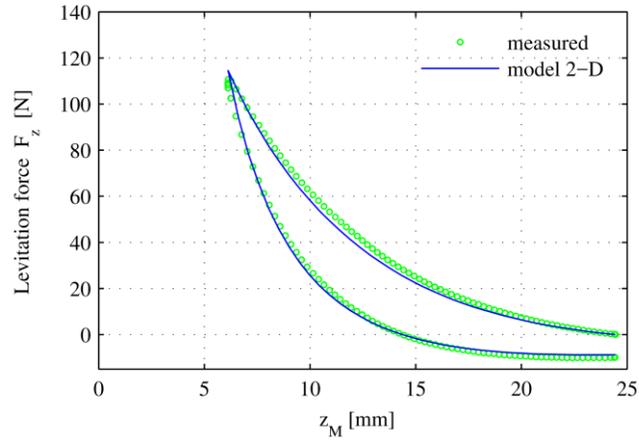

Figure 6 – 2-D model validation: levitation force for the FC25 sequence.

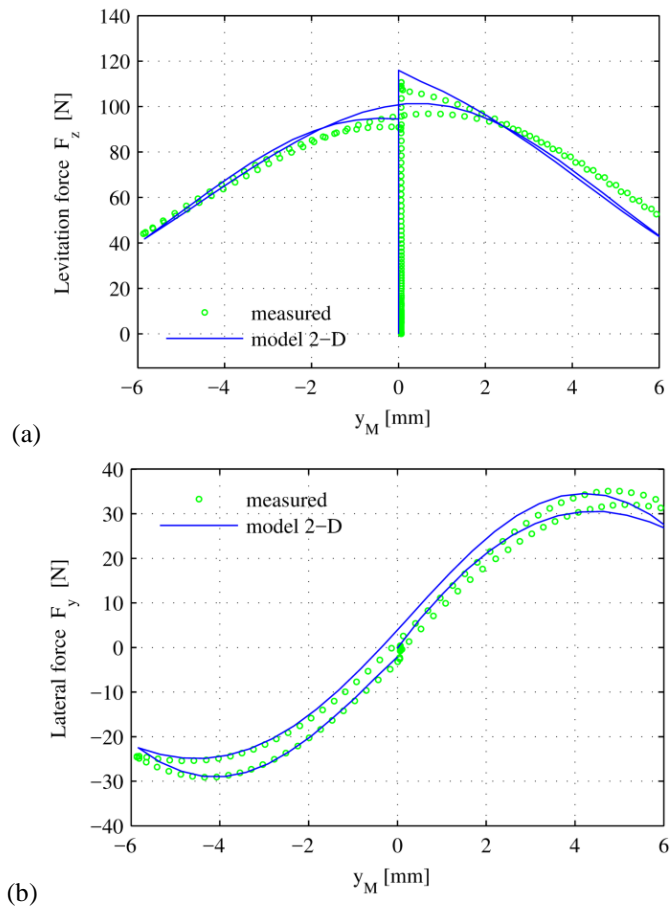

(a)

(b)

Figure 7 – 2-D model validation: (a) levitation force and (b) lateral force for the FC25_LD sequence.

## V. 2D axisymmetric case: axisymmetric SMB

### 1) Geometry

The axisymmetric SMB and the coordinate system adopted in this section are shown in Figure 8. The PM assembly is a cylindrical Nd-Fe-B magnet. The HTS assembly is a cylindrical single domain melt-textured YBCO bulk manufactured by Beijing General Research Institute for Nonferrous Metals. The HTS assembly can only move along the z-direction ($r_M(t) = 0$).

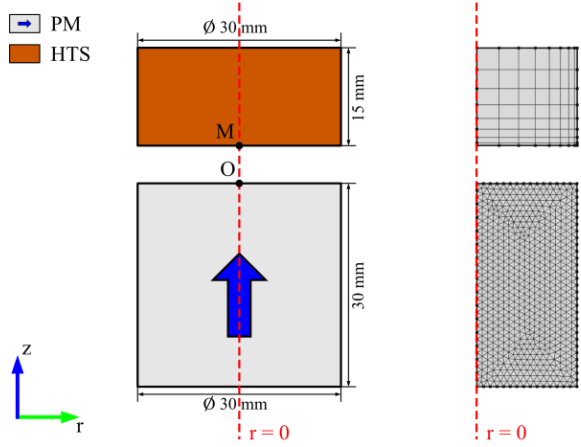

Figure 8 – SMB geometry and mesh for the 2-D axisymmetric case. The point O is located at the center of the PM top surface. The point M is located at the center of the bulk bottom surface. The arrow indicates the PM magnetization direction. The mesh of the air/coolant is not shown.

### 2) Sequences

In this section, we consider three displacement sequences. They are described by the successive positions of $M(r_M, z_M)$ relative to O (in millimeters). The first position of each sequence is the cooling position. The moving speed is 1 mm/s representing a quasistatic process.

- ZFC100: $(r_M, z_M)$ = {(0,100), (0,5), (0,100)}
- FC25: $(r_M, z_M)$ = {(0,25), (0,5), (0,100), (0,5)}
- FC5: $(r_M, z_M)$ = {(0,5), (0,100), (0,5), (0,100)}

### 3) Modeling

Equations (1)-(10) are implemented in COMSOL Multiphysics 4.3a PDE mode application in a 2-D axisymmetric space. More details about such implementation can be found in [76] for example. The HTS assembly is a bulk, thus an isotropic Kim-like model [72] is used to describe the dependence of the critical current density on the magnetic field,

$$J_c(\mathbf{B}) = \frac{J_{c0}}{1 + \frac{|\mathbf{B}|}{B_0}} \tag{15}$$

where $J_{c0}$ and $B_0$ are material parameters. To mesh the HTS bulk, we use the mapped mesh shown in Figure 8 with 8×8 elements distributed following arithmetic sequences in the $rz$-plane. The outer boundary of the HTS assembly model $\Gamma$ is located at 2.5 mm from the HTS bulk, corresponding to half of the minimum levitation gap.

From (9) in 2-D axisymmetric, with the conventions of Figure 8, the expression for $\mathbf{H}_{ext}$ becomes,

$$\mathbf{H}_{ext}(r, z, t) = \mathbf{H}_{PM}(r, z + z_M(t)) \tag{16}$$

where $(0, z_M)$ is the time-dependent position of the HTS assembly relative to O. $\mathbf{H}_{self}$ is obtained by 2-D axisymmetric integration of the Biot-Savart law,

$$H_{self,r}(r, z, t) = \frac{1}{4\pi} \iint_{\Omega_{sc}} \frac{-J_r(r', z', t)\sqrt{m}}{\sqrt{r'r^3}} (z - z') \left[ K(m) - \frac{2 - m}{2(1 - m)} E(m) \right] dr' dz' \tag{17}$$

$$H_{self,z}(r,z,t) = \frac{1}{4\pi} \iint_{\Omega_{sc}} \frac{J_r(r',z',t)\sqrt{m}}{\sqrt{r'r^3}} \, r \left[ K(m) + \frac{m(r+r') - 2r}{2r(1-m)} E(m) \right] dr'dz' \quad (18)$$

$$m = \frac{4rr'}{(r+r')^2 + (z-z')^2} \quad (19)$$

where $\Omega_{sc}$ is the HTS assembly domain, and K and E are the complete elliptic integrals of the first and second kind,

$$K(m) = \int_0^{\pi/2} \frac{d\alpha}{\sqrt{1 - m\sin^2\alpha}} \quad (20)$$

$$E(m) = \int_0^{\pi/2} \sqrt{1 - m\sin^2\alpha} \, d\alpha \quad (21)$$

4) Model calibration

To obtain the remanent flux density $B_r$ of the PM cylinder, we measured the magnetic flux density at several distances above the PM. By a trial-and-error process, we obtained $B_r$ that minimizes the difference between the measured data and the PM assembly model (Figure 9). To obtain $J_{c0}$, it is common practice to use the maximum levitation force obtained for a zero field cooling sequence [34, 56, 26, 50]. Accordingly, the critical current density $J_{c0}$ is set here at $2.4 \cdot 10^8$ A/m$^2$, so that the simulated maximum levitation force during the ZFC100 sequence is equal to the measured value (Figure 10). Alternatively, $J_{c0}$ could have been determined beforehand as done for the 2D case, for example by measuring it by VSM (vibrating sample magnetometer) for a small piece from the bulk as reported in [77]. The other HTS bulk parameters are set to commonly used values. Note that the value of n weakly affects the calculated results if higher than 15. The parameters of the 2-D axisymmetric case are summarized in Table III.

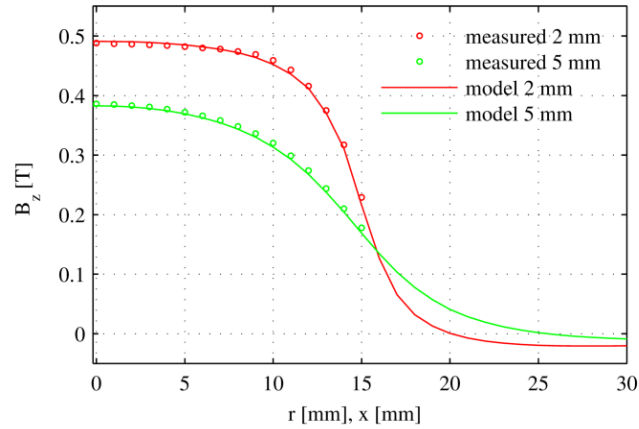

Figure 9 – 2-Daxi / 3-D model calibration: magnetic flux density at 2 mm and 5 mm above the PM.

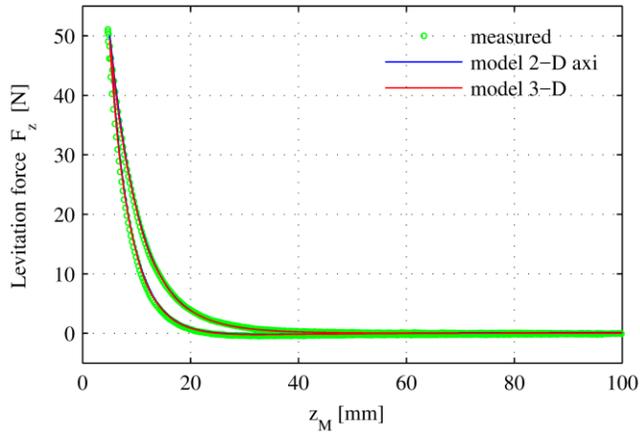

Figure 10 – 2-Daxi / 3-D model calibration: levitation force for the ZFC100 sequence.

TABLE III
PARAMETERS FOR THE 2-D AXISYMMETRIC AND 3-D CASES

| Symbol | Quantity | Value |
| --- | --- | --- |
| $B_r$ | PM remanent flux density | 1.27 T |
| $E_c$ | Critical current criterion | $1 \cdot 10^{-4}$ V/m |
| $J_{c0}$ | HTS parameter | $2.4 \cdot 10^8$ A/m$^2$ |
| $n$ | HTS parameter | 21 [78] |
| $B_0$ | HTS parameter | 0.37 T |
| $\rho_{air}$ | Air resistivity | 1 Ω·m [75] |
| $\mu_0$ | Air/HTS permeability | $4\pi \cdot 10^{-7}$ H/m |

5) Model validation

To validate the 2-D axisymmetric model, we consider the FC25 and FC5 sequences. The force calculated with the 2-D axisymmetric model is in good agreement with the measured force (Figure 11 and Figure 12). This serves as a validation.

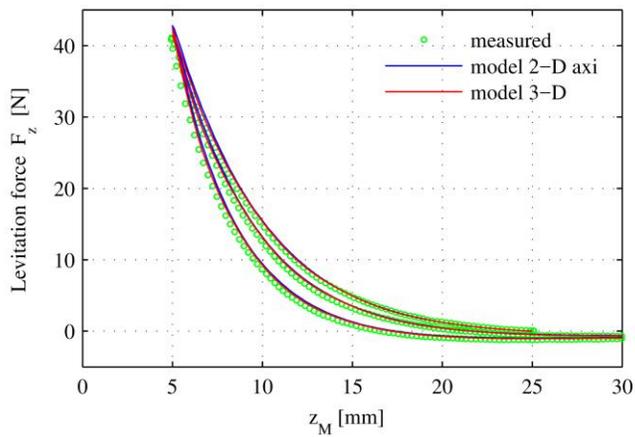

Figure 11 – 2-Daxi / 3-D model validation: levitation force for the FC25 sequence.

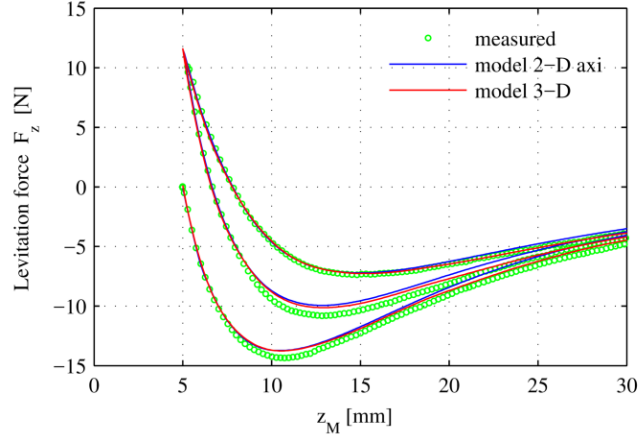

Figure 12 – 2-Daxi / 3-D model validation: Levitation force for the FC5 sequence.

## VI. 3D case
### 1) Geometry

The 3-D SMB and the coordinate system adopted in this section are shown in Figure 13. The SMB is the same as that for the 2-D axisymmetric case but the HTS assembly can now move along any direction.

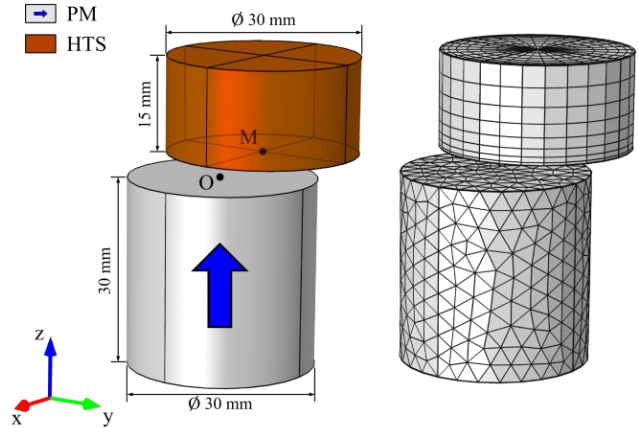

Figure 13 – SMB geometry and mesh for the 3-D case. The point O is located at the center of the PM top surface. The point M is located at the center of the bulk bottom surface. The arrow indicates the PM magnetization direction. The mesh of the air/coolant is not shown.

### 2) Sequences

In this section, we consider six displacement sequences. They are described by the successive positions of $M(x_M, y_M, z_M)$ relative to O (in millimeters). The first position of each sequence is the cooling position. The moving speed is 1 mm/s representing a quasistatic process.

- ZFC100: $(x_M, y_M, z_M)$ = {(0,0,100), (0,0,5), (0,0,100)}
- FC25: $(x_M, y_M, z_M)$ = {(0,0,25), (0,0,5), (0,0,100), (0,0,5)}
- FC5: $(x_M, y_M, z_M)$ = {(0,0,5), (0,0,100), (0,0,5), (0,0,100)}
- ZFC100_Y7.5: $(x_M, y_M, z_M)$ = {(0,7.5,100), (0,7.5,5), (0,7.5,100)}
- ZFC100_Y15: $(x_M, y_M, z_M)$ = {(0,15,100), (0,15,5), (0,15,100)}
- FC25_LD: $(x_M, y_M, z_M)$ = {(0,0,25), (0,0,5), (0,7.5,5), (0,-7.5,5), (0,7.5,5), (0,-7.5,5), (0,7.5,5), (0,0,5)}

The ZFC100, FC25 and FC5 sequences are similar to the 2-D axisymmetric case. The ZFC100_Y7.5 and ZFC100_Y15 sequences are similar to the ZFC100 sequences but the HTS bulk is off-axis.

### 3) Modeling

Equations (1)-(10) are implemented in COMSOL Multiphysics 4.3a PDE mode application in a 3-D space. More details about such implementation can be found in [79] for example. To mesh the HTS bulk, we swept the mesh shown in Figure 8 following a 360° circular path to obtain the hexahedral mesh shown in Figure 13. The outer boundary of the HTS assembly model $\Gamma$ is here again located at 2.5 mm from the HTS bulk, corresponding to half the minimum levitation gap.

From (9) in 3-D, with the conventions of Figure 13, the expression for $\mathbf{H}_{ext}$ becomes,
$$\mathbf{H}_{ext}(y, z, t) = \mathbf{H}_{PM}(x + x_M(t), y + y_M(t), z + z_M(t)) \tag{22}$$

where $(x_M, y_M, z_M)$ is the time-dependent position of the HTS assembly relative to O. $\mathbf{H}_{self}$ is obtained by 3-D integration of the Biot-Savart law,

$$H_{self,x}(x, y, z, t) = \frac{1}{4\pi} \iiint_{\Omega_{sc}} \frac{J_y(z - z') - J_z(y - y')}{\sqrt{(x - x')^2 + (y - y')^2 + (z - z')^2}^3} dx' dy' dz' \tag{23}$$

$$H_{self,y}(x, y, z, t) = \frac{1}{4\pi} \iiint_{\Omega_{sc}} \frac{J_z(x - x') - J_x(z - z')}{\sqrt{(x - x')^2 + (y - y')^2 + (z - z')^2}^3} dx' dy' dz' \tag{24}$$

$$H_{self,z}(x, y, z, t) = \frac{1}{4\pi} \iiint_{\Omega_{sc}} \frac{J_x(y - y') - J_y(x - x')}{\sqrt{(x - x')^2 + (y - y')^2 + (z - z')^2}^3} dx' dy' dz' \tag{25}$$

where $\Omega_{sc}$ is the HTS assembly domain.

### 4) Model calibration

We use the same parameters as that for the 2-D axisymmetric case (Table III).

### 5) Model validation

The 3-D model should be able to reproduce the results obtained with the 2-D axisymmetric model for the ZFC100, FC25 and FC5 sequences. The levitation force calculated with the 3-D model has been added to Figure 10, Figure 11 and Figure 12, showing similar results. To further validate the 3-D model, we consider the ZFC100_Y7.5 and ZFC100_Y15 sequences. The levitation and lateral forces calculated with the 3-D model are in fair agreement with the measured force (Figure 14). The calculated forces are somewhat smaller than the measured ones, but globally the force reduction as a function of the off-axis position is predicted correctly. Similar results have been obtained for a field cooling height of 5 mm (not reported here). Finally, we consider the FC5_LD sequence. The levitation and lateral forces calculated with the 3-D model are plotted together with the measured data in Figure 15. Note the instable behavior of the bearing: when the lateral position increases, the lateral force increases too. Here again, the agreement is fair considering the length of the sequence and the small amplitude of the lateral force. This validates the 3-D model.

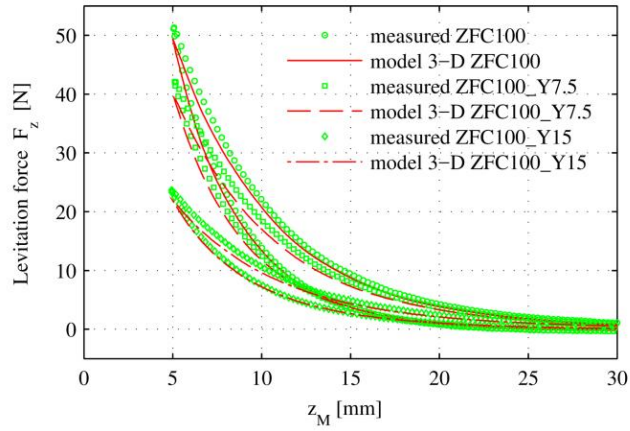

(a)

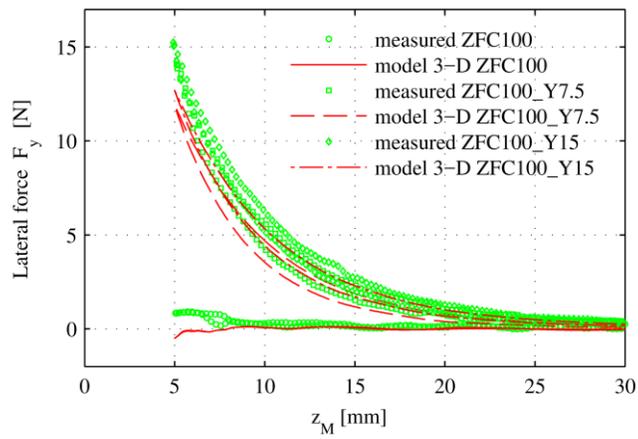

(b)

Figure 14 – 3-D model validation: (a) levitation force and (b) lateral force for the ZFC100, ZFC100_Y7.5 and ZFC100_Y15 sequences.

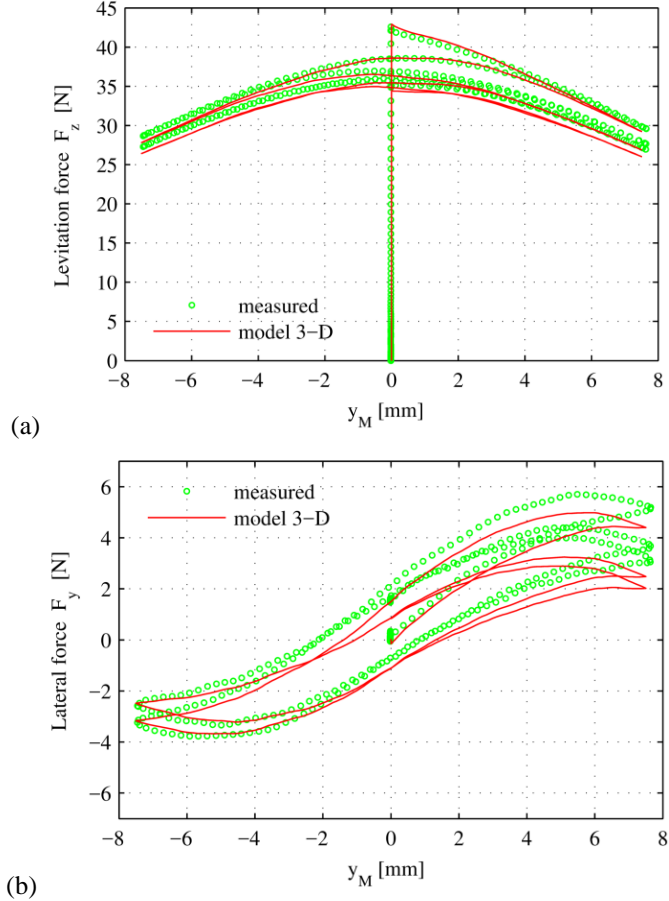

Figure 15 – 3-D model validation: (a) levitation force and (b) lateral force for the FC25_LD sequence.

### VII. Discussion

The test cases considered above have been selected carefully to serve as benchmarks. For the 2-D case, we selected a stack-type SMB for its true 2-D nature. Indeed bulk-type SMB suffer from several factors that make them difficult to be simulated accurately in 2-D. In particular, large bulks with homogeneous properties are difficult to obtain. The end effects and the impact of intragrain currents should then be taken into account [80, 54]. For the 2-D axisymmetric and 3-D cases, we selected a simplistic bulk-type SMB that allows comparison of the results for axial displacement sequences. Finally, we considered on purpose repetitive displacements. This is because simplified models, such as Meissner-limit and frozen-field models, can often estimate the first section of the force loop but generally fail to predict the rest [21, 55].

For the FE discretization, we use linear edge elements [66]. The degrees of freedom of the edge elements being associated with the tangential components along the edges of the elements, it is only possible to impose the tangential component of the field. Nevertheless, in practice a thin layer of air/coolant is sufficient to obtain accurate estimation of the maglev performances as demonstrated in this work.

Melt-textured YBCO bulks have an anisotropic critical current density: it is larger in the ab-plane than along the c-axis [81]. This is the reason why most of previous 3-D SMB models used an anisotropic bulk model. This was either achieved by stacking multiple 2-D layers [43-47, 38, 40], by superimposing two virtual HTS bulks [56] or by considering a tensor of resistivity [49]. As it is still not clear how to model HTS in 3-D to include experimental phenomena such as flux cutting, flux flow and magnetically anisotropic critical current densities [79, 82, 64], we adopted here a simplistic isotropic bulk model. This probably explains the difference between simulation and measurements for the 3-D sequences ZFC100_Y7.5, ZFC100_Y15 and FC25_LD. Indeed, for these sequences the bulk is off-axis and a current is induced along the c-axis. Nevertheless, the present results show that maglev performance of a bulk-type SMB can be reasonably well predicted using a 3-D isotropic bulk model.

To simulate (zero) field cooling, we applied an initial field $\mathbf{H}_0$ according to (3). But because of inherent numerical approximations (mesh, linear elements, etc.), the curl of this field is not perfectly null. In 2-D, this would be equivalent to apply a set of Dirichlet's boundary conditions that doesn't satisfy the Ampère's circuital law on the outer boundary Γ [75]. As a result, some unphysical induced current might flow in the superconducting domain following (4) during (zero) field cooling. A fine mesh was selected here to limit this effect.

The computing time for each model depends on many factors such as: mesh quality, number of time steps, HTS parameters and displacement sequence. All the calculations were performed using Comsol Multiphysics 4.3a [61] and a standard desktop computer (Intel i7-4770S, 3.10 GHz, RAM 8 GB). The state variables were scaled to $10^7$, and the relative and absolute tolerances were set to $10^{-2}$ and $10^{-3}$, respectively. Table IV gives a summary of the computational effort for some sequences. It can seem prohibitive for some applications, in particular when considering complex 3-D SMB geometries. But we used here a rather fine mesh with the goal to obtain good agreements with measurements. Actually coarser meshes can often help to decrease the computing time to few seconds for 2-D cases, without losing too much information [53, 54].

TABLE IV
DEGREE OF FREEDOM, TIME STEPS AND COMPUTING TIME

|  | DOF | Time steps | Computing time [s] |
| --- | --- | --- | --- |
| 2-D ZFC100 | 20270 | 272139 | 392529 |
| 2-Daxi ZFC100 | 539 | 252 | 13038 |
| 3-D ZFC100 | 21068 | 610 | 27777 |

## VIII. Conclusion

We reported here our experience on simulating superconducting magnetic bearing with a commercial finite element software using the H-formulation in 2-D, 2-D axisymmetric and 3-D. The main difficulty is linked to the task of modeling a moving magnet. To address this problem, we chose the approach consisting in modeling the movement via time dependent Dirichlet boundary conditions. It requires (a) only one static solution of the permanent magnet assembly finite element model, and (b) a reduced air/coolant domain around the superconducting material in the HTS assembly model. With a proper calibration procedure, we showed that the proposed model can predict accurately the observed behavior of both stack-tape and bulk-type bearings, for various cooling conditions and various displacement sequences. This comprehensive validation is a necessary step before using such models for designing and optimizing realistic bearings. Besides, the test cases have been selected so that they could be used as a benchmark for other models.

Future efforts could be dedicated to reducing the computing time of such models. For stack-types bearings, the anisotropic homogenization proposed in [62] and extended in [83] is a good alternative. But it should be used with caution, and the first validations proposed in [53, 84] should be extended to other geometries and other test conditions. Another necessary step is the coupling of such models with motion equations, in order to predict the dynamic behavior of the loaded bearing. Indeed, here the relative movement is the input of the simulation but in reality it is a consequence of the efforts exerted on the bearing [85]. Finally, further vetting and refining of the models could help developing and improving lumped parameter SMB models [86, 87], as a mean of drastically speeding up simulations.

**Appendix**

2-D case: iron B-H curve
(B, H) = {(0.0, 0.0), (0.5, 90.0), (1.0, 270.0), (1.1, 318.25), (1.2, 384.50), (1.3, 479.50), (1.3875, 608.562), (1.45, 755.437), (1.5, 939.185), (1.545, 1188.93), (1.575, 1407.93), (1.6275, 2077.31), (1.67375, 3117.93), (1.70225, 3969.37), (1.7275, 4843.66), (1.75825, 6081.34), (1.80875, 8581.09), (1.85, 11066.4), (1.9025, 14985.7), (2.05, 33003.3), (2.15, 59203.3), (2.22625, 93214.9), (2.27, 118884.0), (2.33375, 163558.0), (2.4075, 220788.0), (2.6, 373973.0), (3.0, 692281.0)}

**References**

[1] F.C. Moon, P.Z. Chang, "High-speed rotation of magnets on high Tc superconducting bearings," *Appl. Phys. Lett.*, vol. 56, no. 4, pp. 397–399, Jan. 1990.
[2] B.R. Weinberger, L. Lynds, J.R. Hull, U. Balachandran, "Low friction in high temperature superconductor bearings," *Appl. Phys. Lett.*, vol. 59, no. 9, pp. 1132–1134, 1991.
[3] F.N. Werfel, U. Floegel-Delor, R. Rothfeld, T. Riedel, B. Goebel, D. Wippich, P. Schirrmeister, "Superconductor bearings, flywheels and transportation," *Supercond. Sci. Technol.*, vol. 25, no. 1, pp. 014007-1–014007-16, Jan. 2012.
[4] J.S. Wang, S.Y. Wang, Y.W. Zeng, H.Y. Huang, F. Luo, Z.P. Xu, Q.X. Tang, G.B. Lin, C.F. Zhang, Z.Y. Ren, G.M. Zhao, D.G. Zhu, S.H. Wang, H. Jiang, M. Zhu, C.Y. Deng, P.F. Hu, C.Y. Li, F. Liu, J.S. Lian, X.R. Wang, L.G. Wang, X.M. Shen, X.G. Dong, "The first man-loading high temperature superconducting Maglev test vehicle in the world," *Physica C*, pp. 809–814, 2002.
[5] L. Schultz, O. de Haas, P. Verges, C. Beyer, S. Rohlig, H. Olsen, L. Kuhn, D. Berger, U. Noteboom, U. Funk, "Superconductively levitated transport system - the SupraTrans project," *IEEE Trans. on Applied Superconductivity*, vol. 15, no. 2, pp. 2301-2305, 2005.
[6] L.S. Mattos, E. Rodriguez, F. Costa, G.G. Sotelo, R. de Andrade, R. M. Stephan, "MagLev-Cobra operational tests," *IEEE Trans. on Applied Superconductivity*, vol. 26, no. 3, pp. 1-4, 2016.
[7] W.J. Yang, Z. Wen, Y. Duan, X.D. Chen, M. Qiu, Y. Liu, L.Z. Lin, "Construction and performance of HTS Maglev launch assist test vehicle," *IEEE Trans. Appl. Supercond.*, vol. 16, no. 2, pp. 1108–1111, 2006.
[8] J. Wang, S. Wang, C. Deng, J. Zheng, H. Song, Q. He, and Y. Zeng, "Laboratory-scale high temperature superconducting Maglev launch system," *IEEE Trans. Appl. Supercond.*, vol. 17, no. 2, pp. 2091–2094, 2007.
[9] H. Bornemann, T. Ritter, C. Urban, O. Paitsev, K. Peber, H. Rietschel, "Low friction in a flywheel system with passive superconducting magnetic bearings," *IEEE Trans. Appl. Supercond.*, vol. 2, no. 7-8, pp. 439-447, 1994.
[10] Q.Y. Chen, Z. Xia, K.B. Ma, C.K. McMichael, M. Lamb, R.S. Coolep, P.C. Fopler, W.K. Chu, "Hybrid high Tc superconducting magnetic bearings for flywheel energy storage system," *IEEE Trans. Appl. Supercond.*, vol. 2, no. 7-8, pp. 457-464, 1994.
[11] Y. Miyagawa, H. Kameno, R. Takahata, H. Ueyama, "A 0.5 kWh flywheel energy storage system using a high-Tc superconducting magnetic bearing," *IEEE Trans. Appl. Supercond.*, vol. 9, no. 2, pp. 996–999, 1999.
[12] T. Coombs, A.M. Campbell, R. Storey, R. Weller, "Superconducting magnetic bearings for energy storage flywheels," *IEEE Trans. on Applied Superconductivity*, vol. 9, no. 2, pp. 968-971, 1999.
[13] T. Ichihara, K. Matsunaga, M. Kita, I. Hirabayashi, M. Isono, M. Hirose, K. Yoshii, K. Kurihara, O. Saito, S. Saito, M. Murakami, H. Takabayashi, M. Natsumeda, N. Koshizuka, "Application of superconducting magnetic bearings to a 10 kWh-class flywheel energy storage system," *IEEE Trans. on Applied Superconductivity*, vol. 15, no. 2, pp. 2245-2248, June 2005.
[14] F.N. Werfel, U. Floegel-Delor, T. Riedel, R. Rothfeld, D. Wippich, B. Goebel, G. Reiner, N. Wehlau, "A compact HTS 5 kWh/250 kW flywheel energy storage system," *IEEE Trans. on Applied Superconductivity*, vol. 17, no. 2, pp. 2138-2141, June 2007.
[15] M. Strasik, P.E. Johnson, A.C. Day, J. Mittleider, M.D. Higgins, J. Edwards, J.R. Schindler, K.E. McCrary, C.R. McIver, D. Carlson, J.F. Gonder, J.R. Hull "Design, fabrication, and test of a 5-kWh/100-kW flywheel energy storage utilizing a high-temperature superconducting bearing," *IEEE Trans. on Applied Superconductivity*, vol. 17, no. 2, pp. 2133-2137, June 2007.
[16] S. Mukoyama, K. Nakao, H. Sakamoto, T. Matsuoka, K. Nagashima, M. Ogata, T. Yamashita, Y.

Miyazaki, K. Miyazaki, T. Maeda, H. Shimizu, "Development of superconducting magnetic bearing for 300 kW flywheel energy storage system," *IEEE Trans. on Applied Superconductivity*, vol. 27, no. 4, pp. 1-4, 2017.

[17] K. Nagaya, Y. Kosugi, T. Suzuki, I. Murakami, "Pulse motor with high-temperature superconducting levitation," *IEEE Trans. Appl. Supercond.*, vol. 9, no. 4, pp. 4688-4694, 1999.

[18] J.R. Hull, S. Hanany, T. Matsumura, B. Johnson, T. Jones, "Characterization of a high-temperature superconducting bearing for use in a cosmic microwave background polarimeter," *Supercond. Sci. Technol.*, vol. 18, no. 2, 2005.

[19] T. Matsumura, H. Kataza, S. Utsunomiya, R. Yamamoto, M. Hazumi, N. Katayama, "Design and performance of a prototype polarization modulator rotational system for use in space using a superconducting magnetic bearing", *IEEE Trans. on Applied Superconductivity*, vol. 26, pp. 1-4, 2016.

[20] EBEX Collaboration, "The EBEX balloon borne experiment - optics, receiver, and polarimetry," to be published [Online]. Available: https://scirate.com/arxiv/1703.03847

[21] C. Navau, N. Del-Valle, A. Sanchez, "Macroscopic modeling of magnetization and levitation of hard type-II superconductors: the critical-state model" *IEEE Trans. on Applied Superconductivity*, vol. 23, no. 1, 2013.

[22] C.P. Bean, "Magnetization of hard superconductors," *Physical Review Letters,* vol. 8, no. 6, pp. 250-253, 1962.

[23] J. Rhyner, "Magnetic properties and AC-losses of superconductors with power law current-voltage characteristics," *Physica C*, pp. 292-300, 1993.

[24] C. Hofmann, G. Ries, "Modelling the interactions between magnets and granular high-Tc superconductor material with a finite-element method," *Supercond. Sci. Technol.*, vol. 14, no. 1, pp. 34–40, 2001.

[25] D.H.N Dias, E.S. Motta, G.G. Sotelo, R. de Andrade Jr, R.M. Stephan, L. Kuehn, O. de Haas, L. Schultz, "Simulations and tests of superconducting linear bearings for a maglev prototype," *IEEE Trans. Appl. Supercond.*, vol. 19, no. 3, pp. 2120-2123, 2009.

[26] D.H.N. Dias, E.S. Motta, G.G. Sotelo, R. de Andrade Jr., "Experimental validation of field cooling simulations for linear superconducting magnetic bearings," *Supercond. Sci. Technol.*, vol. 23, pp. 075013+6, 2010.

[27] D.H.N. Dias, G.G. Sotelo, R. de Andrade Jr., "Study of the lateral force behavior in a field cooled superconducting linear bearing," *IEEE Trans. on Applied Superconductivity*, vol. 21, no. 3, 2011.

[28] G.T. Ma, "Considerations on the finite-element simulation of high-temperature superconductors for magnetic levitation purposes," *IEEE Trans. on Applied Superconductivity*, vol. 23, no. 5, 2013.

[29] G.T. Ma, H. Liu, X.T. Li, H. Zhang, Y.Y. Xu, "Numerical simulations of the mutual effect among the superconducting constituents in a levitation system with translational symmetry," *Journal of Applied Physics*, vol. 115, pp. 083908, 2014.

[30] C.Q. Ye, G.T. Ma, J.S. Wang, "Calculation and optimization of high-temperature superconducting levitation by a vector potential method," *IEEE Trans. on Applied Superconductivity*, vol. 26, no. 8, pp. 3603309, 2016.

[31] T. Sugihara, H. Hashizume, K. Miya, "Numerical electromagnetic field analysis of Type-II superconductors," *International Journal of Applied Electromagnetics in Materials*, vol. 2, pp. 183-196, 1991.

[32] N. Takeda, M. Uesaka, K. Miya, "Computation and experiments on the static and dynamic characteristics of high Tc superconducting levitation," *Cryogenics*, vol. 34, no. 9, pp. 745-752, 1994.

[33] Y.D. Chun, Y.H. Kim, J. Lee, J.P. Hong, J.W. Lee, "Finite element analysis of magnetic field in high temperature bulk superconductor," *IEEE Trans. on Applied Superconductivity*, vol. 11, no. 1, 2001.

[34] D. Ruiz-Alonso, T.A. Coombs, A.M. Campbell, "Numerical analysis of high-temperature superconductors with the critical-state model," *IEEE Trans. on Applied Superconductivity*, vol. 14, no. 4, 2004.

[35] L. Wang, H.H. Wang, Q.L. Wang, "Finite element analysis of magnetic levitation force in superconducting magnetic levitation system," *Cryo. & Supercond.*, vol. 34, no. 3, pp. 190-193, 2006. (in Chinese)

[36] G.G. Sotelo, R. de Andrade Jr., A.C. Ferreira, "Test and simulation of superconducting magnetic bearings," *IEEE Trans. Appl. Supercond.*, vol. 19, no. 3, pp. 2083–2086, 2009.

[37] Y.L. Li, J. Fang, M.Z. Guo, L. Xiao, M.H. Zheng, Y.L. Jiao, "ANSYS-based analysis of levitation force in the HTS hybrid magnetic bearings," *Cryo. & Supercond.*, vol. 36, no. 6, pp. 40-44, 2008. (in Chinese)

[38] H. Ueda, S. Azumaya, S. Tsuchiya, A. Ishiyama, "3D electro-magnetic analysis of levitating transporter using bulk superconductor," *IEEE Trans. Appl. Supercond.*, vol. 16, no. 2, pp. 1092–1095, 2006.

[39] A.O. Hauser, "Calculation of superconducting magnetic bearings using a commercial FE-program (ANSYS)," *IEEE Trans. on Magnetics*, vol. 33, no. 2, pp. 1572-1575, 1997.
[40] X.J. Zheng, Y. Yang, "Transition cooling height of high-temperature superconductor levitation system," *IEEE Trans. on Applied Superconductivity*, vol. 17, no. 4, pp. 3862-3866, 2007.
[41] J. Zhang, Y. Zeng, J. Cheng, X. Tang, "Optimization of permanent magnet guideway for HTS maglev vehicle with numerical methods," *IEEE Trans. on Applied Superconductivity*, vol. 18, no. 3, 2008.
[42] X.F. Gou, X.J. Zheng, Y.H. Zhou, "Drift of levitated/suspended body in high-Tc superconducting levitation systems under vibration - Part I: A criterion based on magnetic force-gap relation for gap varying with time", *IEEE Trans. on Applied Superconductivity*, vol. 17, no. 3, pp. 3795-3802, 2007.
[43] M. Uesaka, Y. Yoshida, N. Takeda, K. Miya, "Experimental and numerical analysis of three-dimensional high-Tc superconducting levitation systems," *International Journal of Applied Electromagnetics in Materials*, vol. 4, pp. 13-25, 1993.
[44] Y. Yoshida, M. Uesaka, K. Miya, "Magnetic field and force analysis of high Tc superconductor with flux flow and creep," *IEEE Trans. on Magnetics*, vol. 30, no. 5, pp. 3503-3506, 1994.
[45] M. Tsuchimoto, T. Honma, "Numerical evaluation of levitation force of HTSC flywheel," *IEEE Trans. on Magnetics*, vol. 4, no. 4, pp. 211-215, 1994.
[46] M. Tsuda, H. Lee, Y. Iwasa, "Electromaglev (active-maglev)-magnetic levitation of a superconducting disk with a DC field generated by electromagnets: Part 3. Theoretical results on levitation height and stability," *Cryogenics*, vol. 38, no. 7, pp. 743-756, 1998.
[47] M. Tsuda, H. Lee, S. Noguchi, Y. Iwasa, "Electromaglev (active-maglev)-magnetic levitation of a superconducting disk with a DC field generated by electromagnets: Part 4: theoretical and experimental results on supercurrent distributions in field-cooled YBCO disks," *Cryogenics*, vol. 39, no. 11, pp. 893-903, 1998.
[48] H. Ueda, A. Ishiyama, "Dynamic characteristics and finite element analysis of a magnetic levitation system using a YBCO bulk superconductor," *Supercond. Sci. Technol.*, vol. 17, pp. S170–S175, 2004.
[49] G.T. Ma, J.S. Wang, S.Y. Wang, "3-D modeling of high-Tc superconductor for magnetic levitation/suspension application-Part I: introduction to the method," *IEEE Trans. on Applied Superconductivity*, vol. 20, no. 4, pp. 2219-2227, 2010
[50] G.T. Ma, J.S. Wang, S.Y. Wang, "3-D modeling of high-Tc superconductor for magnetic levitation/suspension application - Part II: validation with experiment," *IEEE Trans. on Applied Superconductivity*, vol. 20, no. 4, pp. 2228-2234, 2010.
[51] S. Pratap, C.S. Hearn, "3-D transient modeling of bulk high-temperature superconducting material in passive magnetic bearing applications" *IEEE Trans. on Applied Superconductivity*, vol. 25, no. 5, pp. 5203910, 2015.
[52] Y. Lu, Y. Qin, "Influence of critical current density on magnetic force of HTSC bulk above PMR with 3D-modeling numerical solutions," *International Journal of Modern Physics B*, vol. 29, nos. 25 & 26, pp. 1542038, 2015.
[53] F. Sass, G.G. Sotelo, R. de Andrade Jr, F. Sirois, "H-formulation for simulating levitation forces acting on HTS bulks and stacks of 2G coated conductors," *Supercond. Sci. Technol.,* vol. 28, no. 125012, 2015.
[54] L. Quéval, G.G. Sotelo, Y. Kharmiz, D.H.N. Dias, F. Sass, V.M.R. Zermeño, R. Gottkehaskamp, "Optimization of the superconducting linear magnetic bearing of a maglev vehicle," *IEEE Trans. on Applied Superconductivity*, vol. 26, no. 3, pp. 3601905, 2016.
[55] A. Patel, S.C. Hopkins, A. Baskys, V. Kalitka, A. Molodyk, B.A. Glowacki, "Magnetic levitation using high temperature superconducting pancake coils as composite bulk cylinders," *Supercond. Sci. Technol.*, vol. 28, pp. 115007, 2015.
[56] Y.Y. Lu, J.S. Wang, S.Y. Wang, J. Zheng, "3D-modeling numerical solutions of electromagnetic behavior of HTSC bulk above permanent magnetic guideway," *J. Supercond. Nov. Magn.*, vol. 21, no. 8, pp. 467-472, 2008.
[57] Z.Q. Yu, G.M. Zhang, Q.Q. Qiu, L. Hu, "Numerical simulation of levitation characteristics of a cylindrical permanent magnet and a high-temperature superconductor based on the 3D finite-element method," *Trans. of China Electrotechnical Society*, vol. 30, no. 13, pp. 32-38, 2015. (in Chinese)
[58] Y.Y Lu, B.J Lu, S.Y. Wang, "The relationship of magnetic stiffness between single and multiple YBCO superconductors over permanent magnet guideway," *J. Low Temp. Phys.*, vol. 164, pp. 279-286, 2011.
[59] Y.Y. Lu, S.J. Zhuang, "Magnetic forces simulation of bulk HTS over permanent magnetic railway with numerical method," *J. Low Temp. Phys.*, vol. 169, pp. 111-121, 2012.
[60] Y.Y. Lu, Q.H. Dang, "Magnetic forces investigation of bulk HTS over permanent magnetic guideway

under different lateral offset with 3D-model numerical method," *Advances in Materials Science and Engineering*, vol. 2012, id. 640497, 2012.
[61] 'COMSOL Multiphysics version 4.3a' Available: www.comsol.com
[62] V.M. Rodriguez-Zermeño, AB. Abrahamsen, N. Mijatovic, B.B. Jensen, M.P. Sørensen, "Calculation of alternating current losses in stacks and coils made of second generation high temperature superconducting tapes for large scale applications," *J. Appl. Phys.,* no. 114, pp. 173901, 2013.
[63] A. Patel, S. Hahn, J. Voccio, A. Baskys, S.C. Hopkins, B.A. Glowacki, "Magnetic levitation using a stack of high temperature superconducting tape annuli," *Supercond. Sci. Technol.*, vol. 30, pp. 024007, 2017.
[64] F. Sirois, F. Grilli, "Potential and limits of numerical modelling for supporting the development of HTS devices," *Supercond. Sci. Technol.*, no. 28, pp. 043002, 2015.
[65] G.T. Ma, H.F. Liu, J.S. Wang, S.Y. Wang, X.C. Li , "3D modeling permanent magnet guideway for high temperature superconducting maglev vehicle application," *Journal of Superconductivity and Novel Magnetism*, vol. 22, no. 8, pp. 841-847, 2009.
[66] R. Brambilla, F. Grilli, L. Martini, "Development of an edge-element model for AC loss computation of high-temperature superconductors," *Supercond. Sci. Technol.*, vol. 20, pp. 16–24, 2007.
[67] V.M. Rodriguez-Zermeño, F. Grilli, F. Sirois, "A full 3D time-dependent electromagnetic model for Roebel cables," *Supercond. Sci. Technol.*, vol. 26, no. 5, pp. 2001, 2013.
[68] H. Huang, J. Zheng, B.T. Zheng, N. Qian, H.T. Li, J. Li, Z.G. Deng, "Correlations between magnetic flux and levitation force of HTS bulk above a permanent magnet guideway," *J. Low Temp. Phys.*, vol. 189, no. 1-2, pp. 42-52, 2017.
[69] C. Navau, A. Sanchez, E. Pardo, D.-X. Chen, "Equilibrium positions due to different cooling processes in superconducting levitation systems," *Supercond. Sci. Technol.,* vol. 17 no. 7, pp. 828–832, 2004.
[70] Y.Y. Lu, Y.J. Qin, Q.H. Dang, J.S. Wang, "Influence of experimental methods on crossing in magnetic force–gap hysteresis curve of HTS maglev system", *Physica C*, vol. 470, no. 22, pp. 1994-1997, 2010.
[71] Z. Hong, A.M. Campbell, T.A. Coombs, "Numerical solution of critical state in superconductivity by finite element software," *Supercond. Sci. Technol.,* vol. 19, no. 12, pp. 1246–1252, 2006.
[72] Y.B. Kim, C.F. Hempstead, A.R. Strnad, "Critical persistent currents in hard superconductors," *Phys. Rev. Lett.*, vol. 9, no. 7, p. 306, 1962.
[73] F. Grilli, F. Sirois, V.M. Rodriguez-Zermeño, M. Vojenčiak, "Self-consistent modeling of the Ic of HTS devices: How accurate do models really need to be?," *IEEE Trans. on Applied Superconductivity*, vol. 24, no. 6, pp. 1-8, 2014.
[74] V.M. Rodriguez-Zermeño, N. Mijatovic, C. Traeholt, T. Zirngibl, E. Seiler, A.B. Abrahamsen, N.F. Pedersen, M.P. Sorensen, "Towards faster FEM simulation of thin film superconductors: a multiscale approach," *IEEE Trans. on Applied Superconductivity*, vol. 21, no. 3, pp. 3273–3276, 2011.
[75] V. Lahtinen, M. Lyly, A. Stenvall, T. Tarhasaari, "Comparison of three eddy current formulations for superconductor hysteresis loss modelling," *Supercond. Sci. Technol.,* vol. 25, no. 11, pp. 115001, 2012.
[76] M. Zhang, J. Kvitkovic, S.V. Pamidi, T.A. Coombs, "Experimental and numerical study of a YBCO pancake coil with a magnetic substrate," *Supercond. Sci. Technol.*, vol. 25, pp. 125020, 2012.
[77] M. Sawamura, M. Tsuchimoto, "Numerical analysis for superconductor in sheet and bulk form," *Japan J. Indust. Appl. Math.*, vol. 17, pp. 199-208, 2000.
[78] M.D. Ainslie, H. Fujishiro, "Modelling of bulk superconductor magnetization," *Supercond. Sci. Technol.*, vol. 28, pp. 053002, 2015.
[79] M. Zhang, T.A. Coombs, "3D modeling of high-Tc superconductors by finite element software," *Supercond. Sci. Technol.*, vol. 25, pp. 01509, 2012.
[80] Z. Deng, M. Izumi, M. Miki, K. Tsuzuki, B. Felder, W. Liu, J. Zheng, S. Wang, J. Wang, U. Floegel-Delor, F.N. Werfel "Trapped flux and levitation properties of multiseeded YBCO bulks for HTS magnetic device applications – Part I: Grain and current features," *IEEE Trans. Appl. Supercond.*, vol. 22, no. 2, pp. 6800110, 2012.
[81] M. Murakami, T. Oyama, H. Fujimoto, S. Gotoh, K. Yamaguchi, Y. Shiohara, N. Koshizuaka, and S. Tanaka, "Melt processing of bulk high Tc superconductors and their application," *IEEE Trans. Magn.*, vol. 27, no. 2, pp. 1479–1486, Mar. 1991.
[82] A. Badía-Majós, C. López, "Electromagnetics close beyond the critical state: thermodynamic prospect," *Supercond. Sci. Technol.*, vol. 25, no. 10, pp. 104004, 2012.
[83] L. Quéval, V.M. Rodriguez-Zermeño, F. Grilli, "Numerical models for AC loss calculation in large-scale applications of HTS coated conductors," *Supercond. Sci. Technol.*, vol. 29, no. 024007, 2016.
[84] K. Liu, W. Yang, G.-T. Ma, L. Quéval, T. Gong, C. Ye, X. Li, Z. Luo, "Experiment and simulation of

superconducting magnetic levitation with REBCO coated conductor stacks," *Supercond. Sci. Technol.*, vol. 31, no. 1, pp. 015013, Dec. 2017.

[85] D.H.N. Dias, G.G. Sotelo, E.F. Rodriguez, R. de Andrade, Jr., R.M. Stephan, "Emulation of a full scale maglev vehicle behavior under operational conditions," *IEEE Trans. on Applied Superconductivity*, vol. 23, no. 3, 2013.

[86] C.S. Hearn, S.B. Pratap, D. Chen, R.G. Longoria, "Lumped-parameter model to describe dynamic translational interaction for high-temperature superconducting bearings," *IEEE Trans. on Applied Superconductivity*, vol. 24, no. 2, pp. 46-53, 2014.

[87] C.S. Hearn, S.B. Pratap, D. Chen, R.G. Longoria, "Dynamic performance of lumped parameter model for superconducting levitation," *IEEE Trans. on Applied Superconductivity*, vol. 26, no. 6, pp. 1-8, 2016.